*Chapter*

# ALTERNATIVE ZEBRA-STRUCTURE MODELS IN SOLAR RADIO EMISSION

## *G.P. Chernov*

Key Laboratory of Solar Activity, National Astronomical Observatories, Chinese Academy of Sciences, Beijing, China

Pushkov Institute of Terrestrial Magnetism, Ionosphere and Radio Wave Propagation, Russian Academy of Sciences (IZMIRAN), Troitsk, Moscow, Russia   e-mail: gchernov@izmiran.ru

## ABSTRACT

In the literature, discussion continues about the nature of the zebra–structure (ZS) in type IV radio bursts, and understanding even the most extended mechanism associated with double plasma resonance has been improved in series of works. Moreover, in the recent work (Benáček, Karlický, Yasnov, 2017) its ineffectiveness was shown under the usually adopted conditions in the radio source. In this case in a number of works we demonstrated the possibility of modeling with whistlers to explain many thin components of ZS stripes, taking into account the effects of scattering whistlers on fast particles. This situation stimulates the search for new mechanisms. For example, earlier we showed the importance of explosive instability, at least for large flares with the ejections of protons. In the system a weakly relativistic beam of protons – nonisothermic plasma, the slow beam mode of the space charge possesses negative energy, and in the triplet slow and fast beam modes and ion-acoustic wave an explosive cascade of harmonics from ionic sound is excited. Electromagnetic waves in the form of ZS stripes appear as a result of the fast protons scattering on these harmonics. Such a mechanism can also be promising for a ZS in radio emission from the pulsar in the Crab nebula.

# 1. INTRODUCTION

In the continuum emission of type IV solar bursts from meter to centimeter wavelength ranges, the following fine structure elements are usually observed: fast pulsations in a wide frequency range; fastest bursts of millisecond duration (spikes); narrow drifting bands in emission and absorption, among which there are fiber bursts with a constant frequency drift and bands with the various drift - the zebra structure (ZS) (Chernov, 2011). Due to the variety of unusual band shapes, the ZS is the most intriguing element presented in the fine structure. Therefore, it has attracted special attention from researchers, and after the first publication on ZS observations in the meter range (Elgaroy, 1959), more than a hundred works have been published, not only on new ZS observations but also theoretical ones.

Over the past 40 years, more than a dozen emission mechanisms have been proposed to explain it. The most popular model is the emission mechanism under double plasma resonance (DPR) (Zheleznyakov and Zlotnik, 1975), in which the upper hybrid frequency $\omega_{UH}$ is compared with an integer number of electron cyclotron harmonics $\omega_{UH} = (\omega^2_{Pe} + \omega^2_{Be})^{1/2} = s\omega_{Be}$, where $\omega_{Pe}$ is the plasma frequency, $\omega_{Be}$ is the cyclotron frequency of electrons, and $s$ is the harmonic number provided that $\omega_{Be} << \omega_{Pe}$. The interaction of the plasma waves with the whistlers can be considered an important alternative mechanism (Chernov, 1976): $l + w \rightarrow t$. Kuijpers (1975) proposed it to explain fiber bursts. However, in a source such as a magnetic trap, whistlers excited by the anomalous Doppler effect can be generated in the form of periodic wave packets. They propagate at a group velocity and determine the frequency drift of the ZS bands, which varies synchronously with the spatial drift of the radio source in the corona. Only the model with whistlers explains a number of fine effects of the dynamics of the ZS bands: the saw-tooth frequency drift, the frequency splitting of stripes, and the superfine structure of stripes in the form of millisecond spikes (Chernov, 2006). A number of other ZS models (Treuman et al., 2011; Ledenev et al., 2006; Laptukhov and Chernov, 2006, 2012; Fomichev et al., 2009; LaBelle et al., 2003; etc.) have not yet been unequivocally recognized. The relevance of ZS research has increased after the discovery of such bands in the radio emission of the pulsar in the Crab nebula under extreme physical parameters peculiar to the pulsars (Zheleznyakov et al., 2012).

The number of works devoted to the improvement of the DPR mechanism continues to grow. A major contribution to this improvement was made by Karlický with co-authors. However, in the last review by the creators of the model (Zheleznykov et al. 2016), this aspect was actually ignored, as well as some perspective offered by new models. Important results were obtained in a recent article by Benáček, Karlický, Yasnov (2017). In this review, we will try to clarify the situation at the moment.

In many solar phenomena, all of the marked elements in the fine structure are present almost simultaneously on a dynamic spectrum, and their emission mechanisms can be different. The use of all available optical and X-ray data often helps in the flare analysis. Thus, in the phenomenon that occurred on April 11, 2013, the use of simultaneous images in several extreme ultraviolet lines from the Solar Dynamic Observatory/Atmospheric Imaging Assembly (SDO/AIA) helped researchers to understand the repeated change in the sign of the circular polarization of the radio emission, when each new flare brightening occurred over regions with different magnetic polarities, and the ordinary wave mode remained (Chernov et al., 2016). Additional data in the case of the hard X-ray emission (RHESSI) made it possible to construct a probable radio source scheme within the standard model of a flare with the magnetic reconnection, where the pulsations of radio emission in the 2.6–3.4 GHz band were associated with the upward acceleration of fast particles from the current sheet, and the ZS emission at lower frequencies was associated with the capture of particles accelerated downward, into a flare loop.

However, in the following phenomenon on June 21, 2013, a group of fast pulsations also developed around 3 GHz, but there was no ZS. In the phenomenon on December 1, 2004, on the contrary, from the very beginning of the phenomenon, there were different fiber bursts, spikes, pulsations, and ZS almost simultaneously on the spectrum in the decimeter range of 1.1–1.34 GHz.

Thus, the main hierarchy of the fine structure can be related with the dynamics of the flare process. The statistical analysis of the ZS is complicated by a wide variety of phenomena (Tan et al., 2014). In the absence of the high-resolution positional observations of radio sources, it is first important to understand the causes of the sequential appearance of individual elements in the fine structure, and it is even more important to understand their simultaneous appearance, since no such analysis of phenomena has been conducted so far.

First, we show some complex examples of observations of the ZS in different frequency ranges, which are difficult to interpret within the framework of DPR mechanism. Then we pass directly to theoretical problems.

## 2. New observations of a complex zebra-pattern

In this section, we intend to show complex examples of ZS stripes that are really difficult to explain within the framework of the DPR mechanism.

Numerous ropes of the fiber-bursts were observed for three minutes in the flare continuum as a part of the developed ZS after the group of strong type III + V bursts shown in the IZMIRAN spectrum in the meter range in Figure 1. This event shows that a mechanism associated with the ZS must include the possibility of explaining the complex rope-like fibers and strange fibers in the absorption, submerged in the developed ZS.

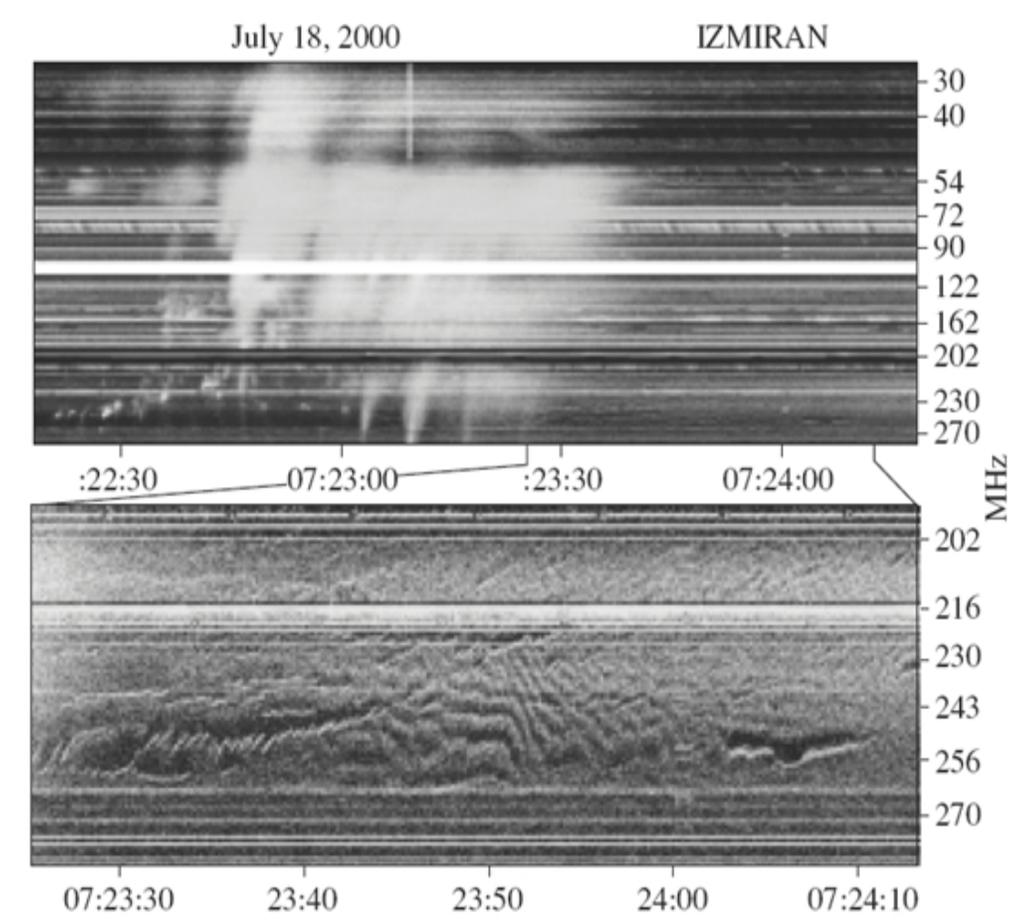

Figure 1. IZMIRAN dynamic spectrum (25–270 MHz) for the beginning of the July 18, 2000 event (upper panel). The lower panel shows a magnified fragment of the rope of fibers in the developed ZS. (A part of Figure 9 from Chernov, 2008).

Figure 2 shows a pulsating character of the ZS, with simultaneous radio fibers (fiber-bursts) at the low frequency part of the decimeter range, the oscillating frequency drift, a transformation of the ZS stripes into the fiber bursts and inversely and other fine structure, e.g. the frequency splitting of the ZS stripes.

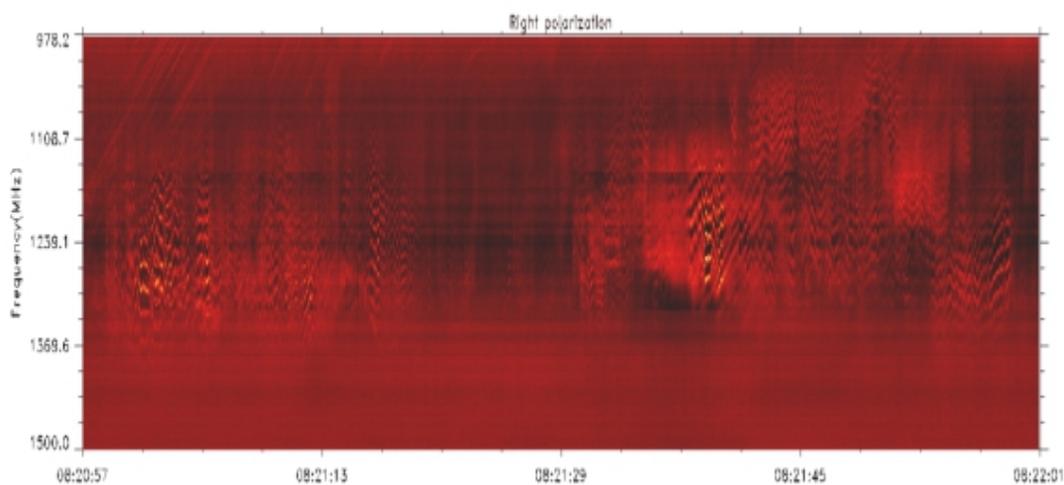

Figure 2. Zebra- structure in the pulsating regime in the event on August 1, 2010 in the decimeter range of 1000 – 1500 MHz observed by the spectro-polarimeter of Yunnan Observatory (YNAO, Kunming, China). The radio emission is weakly polarized. (Figure 1 from Chernov, 2017, Arxiv.org/abs/1704.02528).

Spectra in Figure 3 in the decimeter range on December 1, 2004 show practically simultaneous presence of spikes, fiber bursts and ZS stripes. What is more, they transform into each other: from a chaos of spikes into fiber bursts and ZS and back, and besides, the spikes look like primary radiation elements.

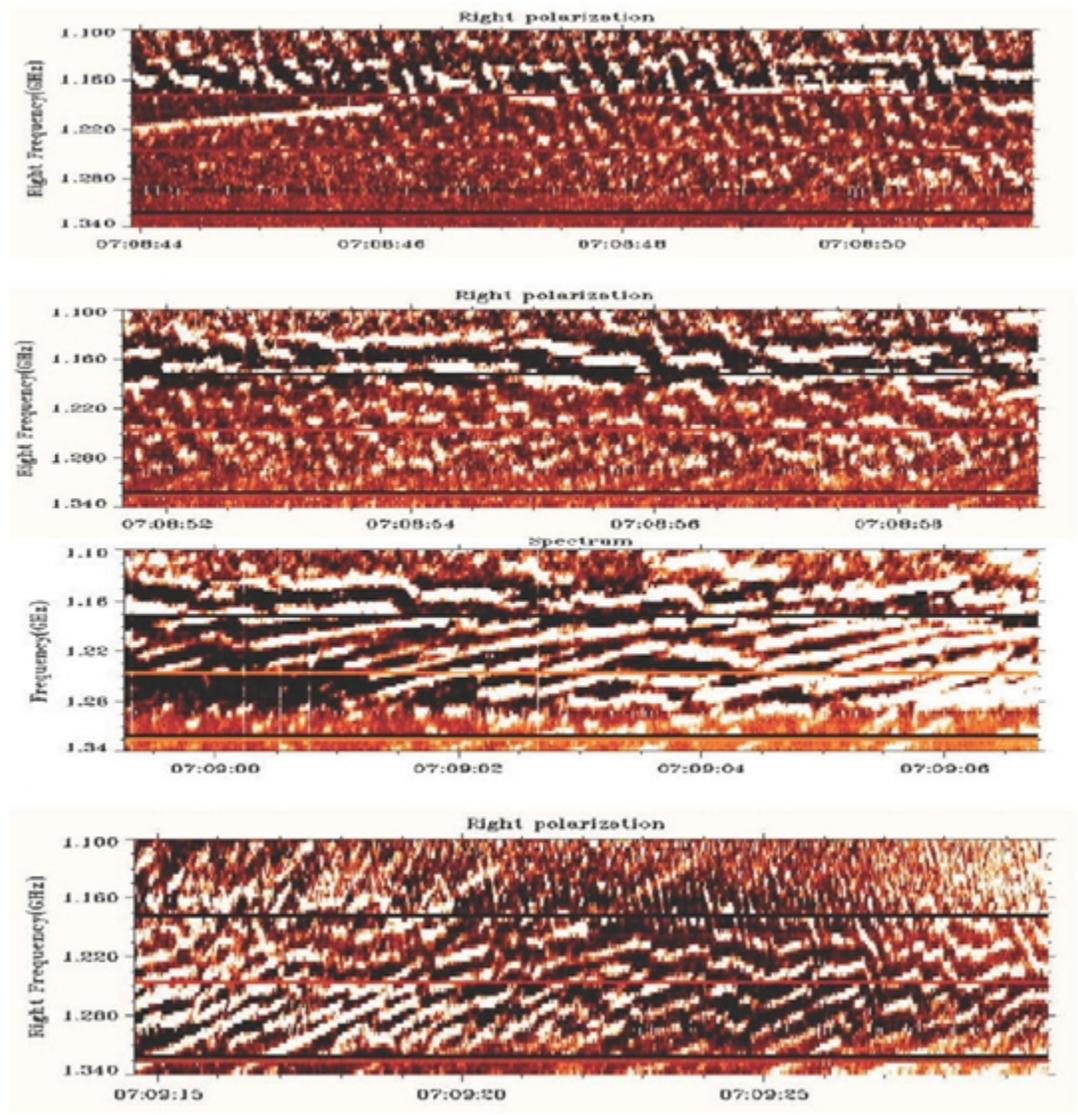

Figure 3. Gradual transformation of spikes and fiber bursts into the zebra stripes (and back) and its evolution within 46 s. (Figure 15 from Chernov et al. 2017).

In Figure 4 the richness of fine structures is shown around 7000 GHz: zebra stripes and radio fibers of different scales and frequency drift, as well as the superfine structure apparent as millisecond spikes.

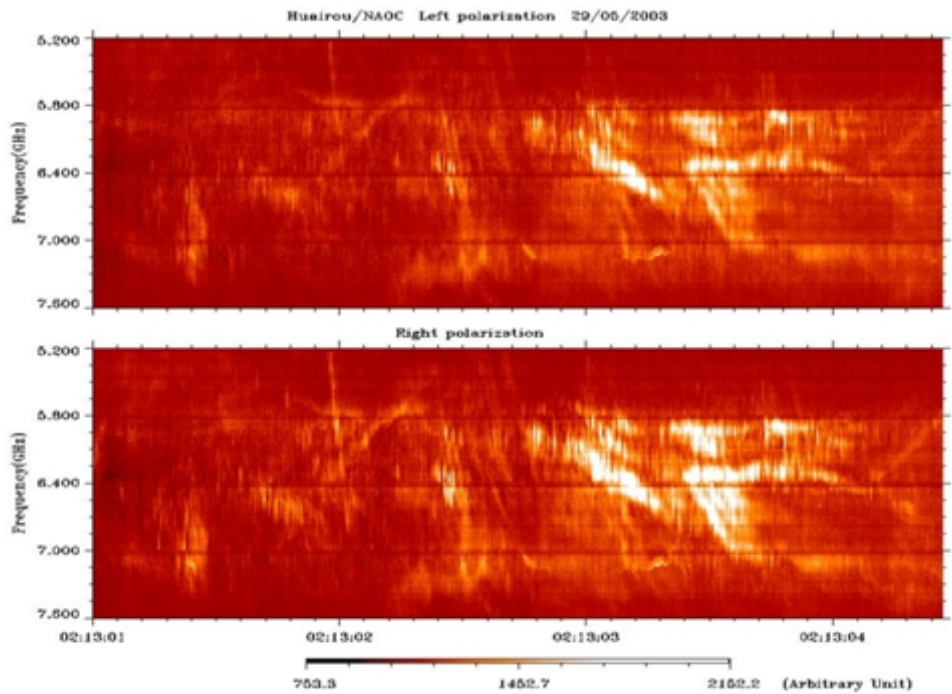

Figure 4. Complex fiber bursts and ZS with the superfine structure in the microwave range of 5.2–7.6 GHz from the event on May 29, 2003 (a fragment of Figure 2 from Chernov et al. 2012).

It is also possible to add the spectra from the April 21, 2002 event with the 34 ZS stripes in the range of 2.6-3.8 GHz (Figures 5 and 6 in Chernov et al. 2005). Some other unusual forms of the ZS:

- Radio fiber bursts are sometimes imposed on the ZS, or limited it by the LF edge in the decimeter range (or from HF edge in the microwave range);

- A ZS appears in the pulsing mode, with chaotic instantaneous columns (almost without their drift) showing random duration from 0.1 s to 6 s;

- A wavelike or saw-tooth frequency drift of the ZS stripes in the columns.

- From another event we had many examples of the frequency splitting of ZS stripes and there was a synchronous change in the frequency drift of ZS stripes with a spatial drift in their radio sources. Perhaps, this was the main property the ZS.

## 3. Problems with the DPR mechanism

After the last big review devoted to the DPR mechanism by Zheleznyakov et al. (2016), we do not need a detailed description of it here. Really, the most promising (and the most cited) mechanism is based on the DPR instability of the plasma together with the loss-cone type of electron distribution function. In this model, firstly the upper-hybrid waves are generated and then these waves are transformed to electromagnetic (radio) waves with the same or double frequency of the upper-hybrid waves. The beautiful theory, presented by Zheleznyakov & Zlotnik (1975) and supported by Winglee & Dulk (1986) became almost classical. The basic condition of instability (the existence of DPR levels, if the scale of the magnetic field heights $L_B$ is much smaller than the scale of the plasma density heights $L_N$) seems obvious. However, the use of hypothetical schemes of dependencies of the cyclotron frequency harmonics $s\omega_{Be}$ and the plasma frequency $\omega_{Pe}$ as functions of the coordinate $x$ without digital scales on the axes, immediately raised questions, both in the first papers and in many subsequent ones, including the last review by Zheleznyakov et al. (2016) (see e.g. Figure 5 as a part of Figure 2 from the last review).

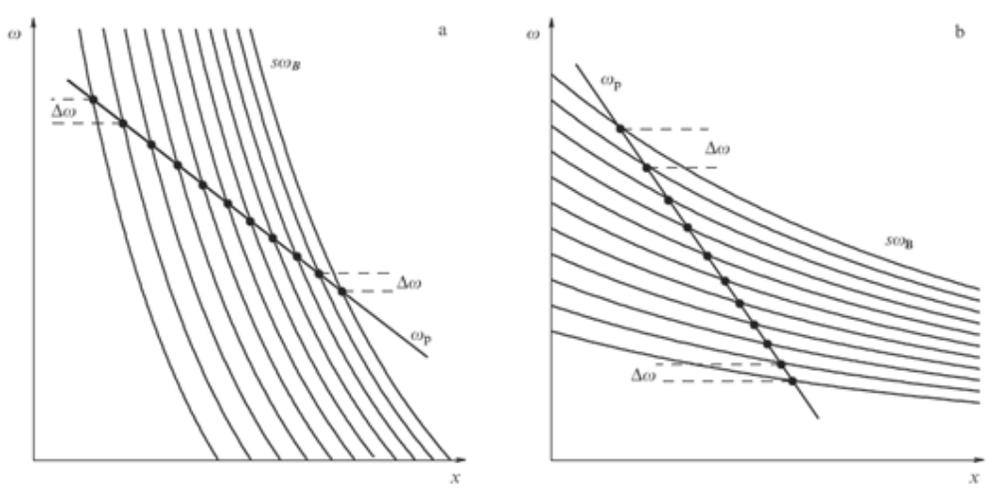

Figure 5. Cyclotron frequency harmonics $s\omega_{Be}$ and the plasma frequency $\omega_{Pe}$ as functions of the coordinate $x$: (a) for the same signs of $L_B$ and $L_N$ and for $|L_B| < |L_N|$; (b) the same as in panel (a) but for $|L_B| > |L_N|$ (a part of Figure 2 from Zheleznyakov et al. (2016)).

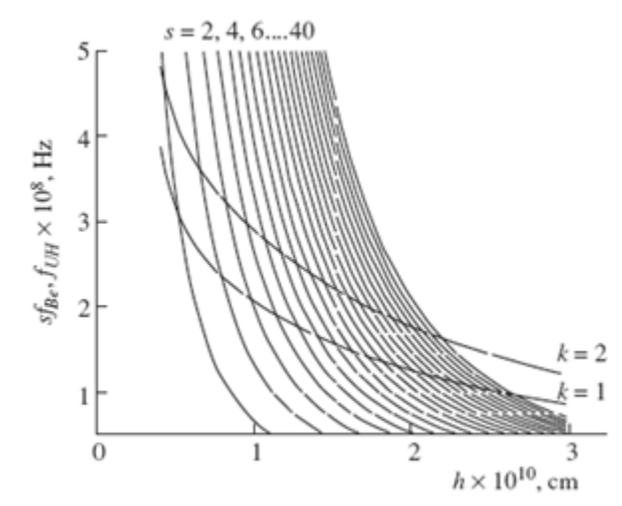

Figure 6. Upper hybrid frequency $f_{UH}$ and electron gyrofrequency harmonics $sf_{Be}$ in the solar corona as functions of height $h$ in the Newkirk model ($k = 1$) and the double Newkirk model ($k = 2$) of the electron density profile ($n = 16.52 \times 10^4 \times 10^{4.32/h}$ cm$^{-3}$, where $h$ is the distance from the center of the Sun) for a dipole magnetic field with $B$ = 2500 G at the photosphere level. (Figure 3 from Chernov, 2005).

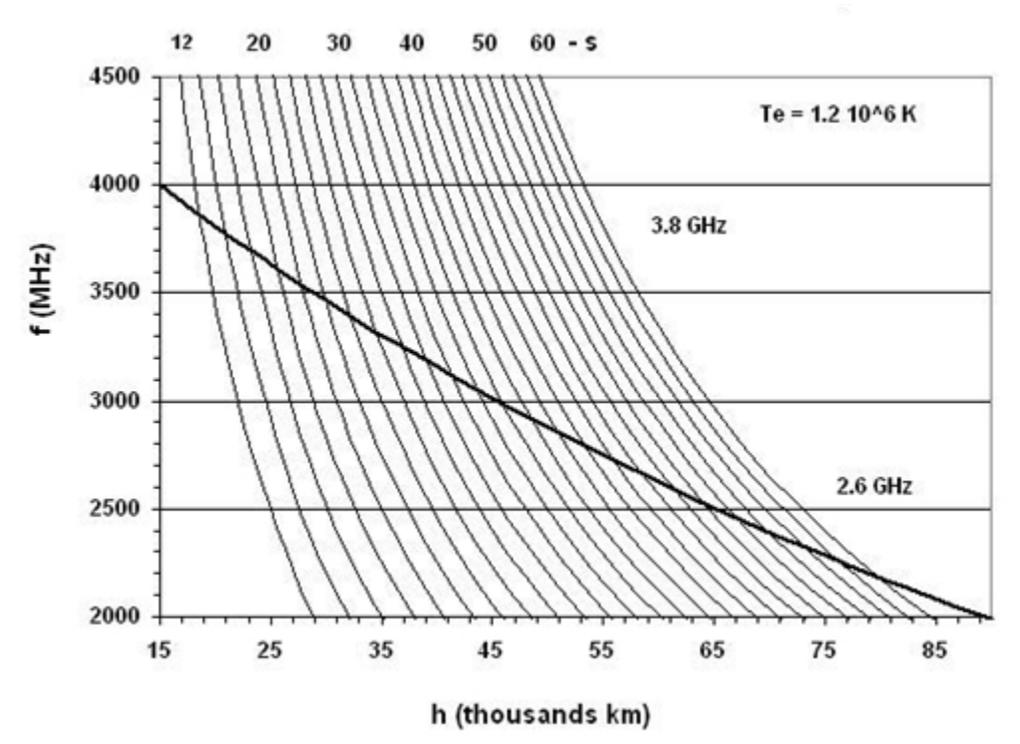

Figure 7. Altitude dependence of the plasma frequency in accordance with the barometric law (heavy line) and altitude profiles of the electron cyclotron harmonics $s$ (light lines) in the solar corona. For the electron temperature $T_e$ = 1.2 × $10^6$ K and initial frequency $f_{P0}$ = 3800 MHz at an altitude of $h_{B0}$ = 20 000 km between the plasma layers associated with 2600 - 3800 MHz, 34 DPR levels are formed (from Laptuhov & Chernov, 2009).

But if we use real known analytical expressions for the electron density and magnetic field strength, the obtained DPR levels do not correspond to observations both in terms of their number and in the height distribution in the corona. As shown in Figure 6 for the event on October 25 1994 in the meter range, instead of the observed eighteen stripes in the 135-170 MHz frequency range, we will obtain only ten stripes with harmonic numbers $s$ varying from 10 to 20, which are extended in the corona to the too high altitudes of 200 thousand km. Moreover, this model predicts a sharp increase in the frequency separation between the stripes (from 2.5 to 7 MHz) rather than an actually observed gradual increase from 1.7 to 2.2 MHz.

For the microwave range we consider the possibility of simultaneous excitation of waves at 34 DPR levels in the corona, assuming that the plasma density depends on the altitude by the conventional barometric formula $f_P = f_{P0} \exp[-(h - h_{B0})/10^4 T]$ and the magnetic field by the formula derived in Dulk & McLean (1978) from the radio data, $B$ = 0.5

*(h/Rs)* $^{-1.5}$ where *Rs* is the Sun's radius, and we obtain that the 34 DPR levels extend in the corona up to too high altitudes of 80 000 km, which, according to current knowledge,

correspond to the plasma frequency ∼250 MHz. We calculated the DPR levels shown in

Figure 7 by using a barometric formula with the commonly accepted coronal plasma parameters: the electron temperature $T_e = 1.2 \times 10^6$ K and the initial plasma frequency $f_{P0} = 3800$ MHz at the altitude $h_{B0} = 20000$ km. If we use a dipole dependence of the magnetic field for the cyclotron harmonics, then the DPR resonances at harmonics with $s > 50$ will occur at altitudes higher than 100000 km. Thus, the simultaneous excitation of waves at 34 levels in the corona is impossible for any realistic profile of the plasma density and magnetic field (if we do not assume the local density and magnetic field scale heights to be smaller in terms of order of magnitude).

At last, observations by Chen et al. (2011) in the decimeter range with the high frequency and spatial resolution provide direct evidence that different ZS stripes are generated in a magnetic flux tube in spatially separated sources. Now, this paper is cited as the decisive observational argument favoring the ZS origin due to the DPR effect in the solar corona. However, the main problem is decided by the usual method: by selecting the relationship of the height scales and their numerical values. The most probable relationship is selected ($L_n/L_B = 4.4$) so that the DPR levels for the observed ZS stripes hit into the observed interval of heights. The arbitrary exponential dependences of density and magnetic field with height in the corona were selected. Then the observed size of source and selected relationship $L_n/L_B = 4.4$ give the values of height scales. Such a procedure is still far from the direct observations of positions of the sources for each stripe. Furthermore, the authors erroneously rejected model with whistlers, as this has already been noted in the review Chernov (2016).

Karlický and Yasnov (2015) confirmed that in models with smaller density gradients, e.g. in those with the barometric density profile, the microwave zebras cannot be produced owing to the strong bremsstrahlung and cyclotron absorption affecting the non-thermal electrons. And they showed that microwave zebra stripes are preferentially generated in regions with steep gradients of the plasma density, for example in the transition region, based on the the previous results of the authors (Yasnov and Karlický, 2015) using the density model by Selhorst, et al. (2008).

The appearance of a number of works on improvement of the DPR mechanism is obviously connected with some doubts in the results of calculations from the first works (Zheleznykov and Zlotnik, 1975; Zlotnik et al.2003), although they are confirmed once more in the last review (Zheleznykov et al. 2016).

In the works Chernov et al. (2005); Chernov (2006); Chernov (2011) a specific doubt is expressed relative to a narrow peak in the hybrid band ($\delta\omega / \omega_B \sim 2.5 \times 10^{-4}$) in calculations of the growth rate of the plasma waves at the upper hybrid frequency $\gamma_{UH}$. The reason is that at the estimation of $\Delta k_\perp / k_\perp$ and $k_{||}^{opt}$ (the expression B.15 in Zlotnik et al. (2003) and formula (22) in Zheleznykov et al. (2016)) the velocity dispersion $\Delta \upsilon_\perp / \upsilon_\perp$ was missed as an infinitesimal quantity.

The DPR resonance condition in Zheleznyakov and Zlotnik (1975) is only valid in the zero-temperature limit. In the recent paper by Benáček, Karlický, Yasnov (2017) they analyzed effects of temperatures of phone plasma and hot electrons on zebra generation processes using the resonance condition in its general form, taking into account the relativistic factor.

The obtained values of growth rate behavior differs from those in previous studies (Zheleznyakov & Zlotnik 1975; Winglee & Dulk 1986). The difference is caused by used relativistic corrections and mainly by selection of maximal growth-rates in $k$-maps. The values of the parallel components $k_{||}^{opt}$ substantially increase, and the values of the perpendicular component $k_{\perp}$ decrease, with the retention of the inequality $k_{||}^{opt} << k_{\perp}$. The maximal growth-rates are presented in Benáček, Karlický, Yasnov (2017) in Figure 4, and it's upper part is shown here in Figure 8. It shows that for a relatively low temperature of hot electrons (the thermal velocity of hot electrons $\upsilon_t = 0.1c$) the dependence of the growth-rate vs. the ratio of electron plasma and electron cyclotron frequencies expresses distinct peaks (for three first harmonics) and with increasing this temperature ($\upsilon_t = 0.2c$) these peaks are smoothed.

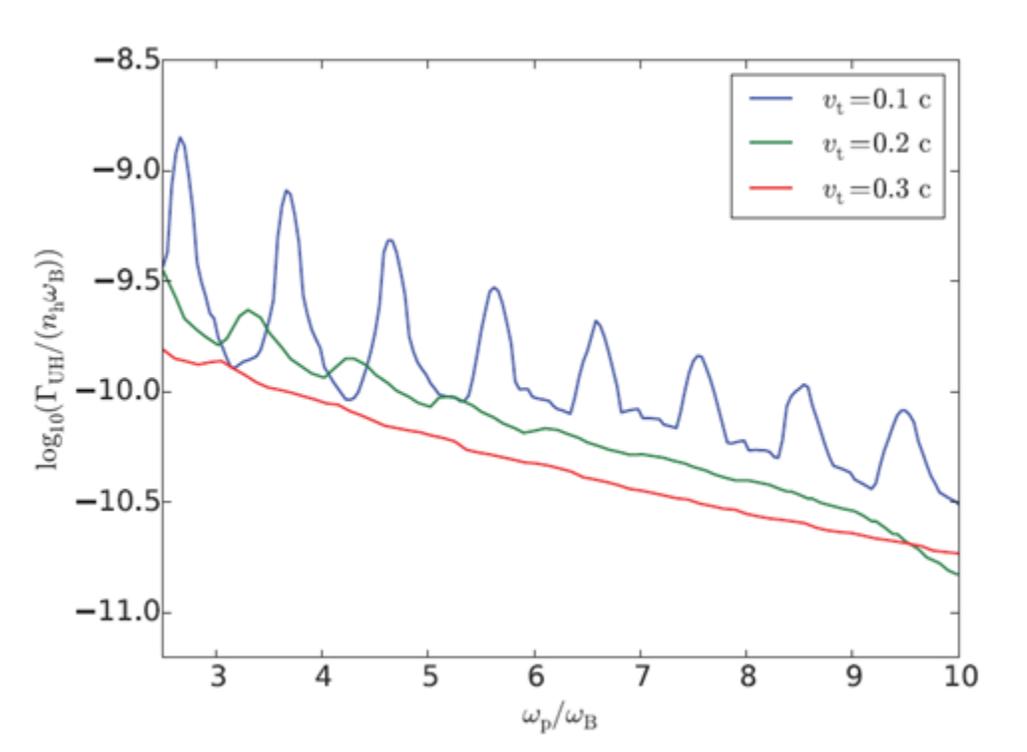

Figure 8. Growth-rates that depend on $r_{pB} = \omega_p / \omega_B$ with the fixed background temperature $\upsilon_{tb} = 0.018$ c. (A part of Figure 4 from Benáček, et al. 2017).

Moreover, as seen in Figure 8, also the relative bandwidth of the growth rate maxima increases with the increase of the harmonic $s$. Such a behavior differs crucially from qualitative estimations shown in Figure B1 in Zlotnik et al. (2003) and Figure 2 in Zheleznykov et al. (2016), showing the ineffectiveness of the DRR mechanism for the ZS. These new calculations are obtained not so much due to complexity and volume of calculations, but rather due to the detailed account of all parameters without simplifying assumptions. In this paper (Benáček, et al. 2017) the DPR instability was studied analytically.

In the recent paper by Benáček and Karlický (2017) these results were verified using a 3D particle-in-cell (PIC) relativistic model in two versions (multi-mode and specific-mode models). The authors computed not only the growth rates of the DPR instability, but also the dynamics of the distribution function and saturation energies of the generated upper-hybrid waves. The loss-cones of the distributions are fulfilled step by step by electrons and thus the distributions become closer to the Maxwellian distribution. These changes are due to an increasing energy level of the upper-hybrid waves. In Figure 3 of Benáček and Karlický (2017), a very important result was shown that the resonant ellipses shifts with the increase of $k_{\parallel}$ and $k\perp /c$.

They found a very good agreement between numerical and analytical results. Figure 6 in Benáček and Karlický (2017) almost repeats Figure 4 (growth-rates dependence) in Benáček, et al. (2017), in more details. In the PIC model these results were obtained by quite different methods. In the PIC model they simulate the plasma as a cloud of particles with an enormous number of electrons and protons, which self-consistently interact through Maxwell equations. They computed the saturation energies of the generated upper hybrid waves, which is beyond the possibility of the analytical theory. They found that the saturation energy decreases with increasing $s$ and this decrease is exponential.

Earlier, Kuznetsov & Tsap (2007) showed that stripes of a zebra pattern become more pronounced in the DPR model with an increase in the loss-cone opening angle and the power-law spectral index. However, the above mentioned problems demonstrate the remaining impossibility of realizing a sufficient number of DPR levels in any analytical density model in the corona.

## 4. Brief discussion and conclusion

Let us refine after a review of the above Figures 1-4, what any model should explain:

- the radio fibers are sometimes imposed on the ZS, or limit it by the LF edge in the decimeter range (or from HF edge in the microwave range);

- the ZS appears in the pulsing mode, by chaotic instantaneous columns (almost without their drift) with random duration from 0.1 s to 6 s;

- a wavelike or saw-tooth frequency drift of stripes of the ZS in columns;

- the frequency separation grows with frequency (as usual);

- superfine structure of ZS stripes and continuum.

From the other events we had many examples of the frequency splitting of the ZS stripes and there was a synchronous change in the frequency drift of the ZS stripes with the spatial drift of their radio sources. Perhaps, this was the main property the ZS.

In the DPR model, all the changes of ZS stripes are usually associated with changes in the magnetic field and with propagating fast magneto-acoustic waves. In a very complicated case the simultaneous presence of fast particles in a radio source with several different distribution functions was proposed (Zheleznyakov et al. 2016).

At the same time, in the whistler model all the aforementioned properties of ZS stripes mentioned above were explained by real physical processes occurring during the coalescence of the Langmuir waves with the whistler waves (Chernov, 2006).

First of all, a radio source of the ZS is usually located in magnetic islands after CME ejections. Therefore, the close connection of the ZS with the fiber bursts is simply explained by the acceleration of fast particles in magnetic reconnection regions in the lower part and in the upper part of magnetic islands.

Let us recall that a wavelike or saw-tooth frequency drift of the stripes was explained by the switching of whistler instability from the normal Doppler cyclotron resonance into the anomalous one, qualitatively shown in Figure 9 for the distribution function $F(V_\parallel, V_\perp)$ (Chernov, 1990).

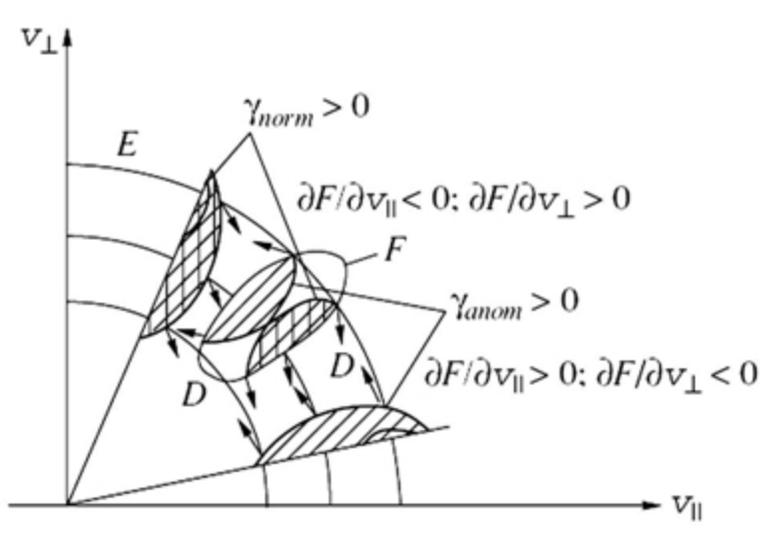

Figure 9. Switching instability of the whistlers from the normal Doppler effect to anomalous one, according to Gendrin, 1981; $F$ – levels of a distribution function, $E$ – levels of equal energy, $D$ – electron diffusion directions. ( From Chernov, 1990).

In the model with the whistlers, many components of the dynamics of the ZS stripes are explained by the quasi-linear effects of the diffusion of the whistlers on fast particles. The smooth switching of the predominant contribution from the anomalous to normal Doppler resonances (and inversely) occurs in accordance with the sign of an operator $\Lambda$ in the growth rate expression (Gendrin, 1981; Bespalov and Trahtengerts, 1980):

$$\Lambda = \frac{s\omega_B}{\omega V_\perp} \frac{\partial}{\partial V_\perp} + \frac{k_\parallel}{\omega} \frac{\partial}{\partial V_\parallel} \mid \quad V_\parallel = (\omega - s\omega_B) / k_\parallel$$

It is known that in the normal Doppler effect, particles and waves propagate in the same direction, but in the anomalous Doppler effect, they propagate in opposite directions. This effect provides a smooth change in the whistler propagation direction and, consequently, a smooth change in the frequency drift of stripes. This effect is called a fan instability in the tokamak plasma (Parail and Pogutse, 1981).

Such switching should lead to a synchronous change in the frequency drift of stripes and spatial drift of the radio source. New injections of fast particles cause a sharp change in the frequency drift of stripes in the instantaneous ZS columns. Low frequency absorptions (black stripes of ZS) are explained by weakening of the plasma wave instability due to the diffusion of fast particles by whistlers.

The superfine structure is generated by a pulsating regime of the whistler instability with ion-sound waves (Chernov et al. 2003). Rope-like chains of the fiber bursts are explained by the periodic whistler instability between two fast shock fronts in a magnetic reconnection region (Chernov, 2006). In the whistler model zebra stripes can be converted into fiber bursts and spikes (and back), as is shown in Figure 3.

Yasnov et al. (2017) shown that in the DPR model high brightness temperature of electromagnetic waves with $T_b \geq 10^{15}$ K can be obtained in a source with a steep density gradient in the density model of Selhorst et al. (2008) in the transition region. Simultaneously they estimated a very high level of saturation energy for the upper-hybrid waves, $W_{UH} = 1.6 \times 10^{-3} E_{h,kin}$ (kinetic energy of hot electrons). In this connection, it is strange that Zheleznyakov et al. (2016) repeat their argument against the whistler model using the Manley-Rowe relation for the brightness temperature of electromagnetic radiation in the result of the coupling of plasma and whistler waves, although the error in this argument was noted several times (e.g. in Chernov et al. 2015). Zheleznyakov et al. (2016) states that since $\omega_w << \omega_l$, in the denominator, only the first term remains and $Tb$ depends only on $T_l$, and $T_b \sim T_l$, i.e. the process does not depend on the level of whistler energy. However, in several papers it was shown that the second term $\omega_w T_l$ should be $>> \omega_l T_w$ due to $T_l >> T_w$. Therefore the value of $Tb$ in the process $l+w \rightarrow t$ depends mainly on $T_w$.

## 4.1. The new alternative mechanism of the ZP due to development of explosive instability in the system beam – plasma

Observations of the ZPs during powerful flares make it possible to assume that the particle acceleration to relativistic velocities and excitation of different wave modes occur in the radio source. Therefore, probably, other possible interactions of waves and particles should be taken into account. For example, Fomichev & Fainshtein (1981) proposed the decay instability of whistlers to the harmonics of the ion sound, which have weak spatial dispersion and small damping at frequencies, much less than the ion Langmuir frequency $\omega_{0i}$. In Fomichev et al. (2009) an alternative mechanism of the ZPs is discussed, due to the development of explosive instability in the system, which is a weakly relativistic beam with nonisothermic plasma. The explosive instability appears in the nonequilibrium system, where there are waves of negative energy (Kadomtsev et al. 1964), moreover, in the resonance triplet the wave with the highest frequency $\omega_3$ must possess negative energy, and the two lowest waves ($\omega_{1,2}$) have positive energy. Fomichev et al. (2009) have shown that the mechanism of generating the ion-acoustic "saw", as a result of the development of explosive instability in the system with a weakly relativistic flow of protons and strongly nonisothermic plasma is more effective in the sense of energy.

The number of harmonics of ionic sound $n$ is determined by two factors:

1) the dispersion of ion sound must be sufficiently small;

2) the number of effective collisions $\nu_{ef} \ll \omega$ ($\omega$ – the angular frequency of sound), i.e.,

the $n$-th harmonic of sound must weakly attenuate. The beam ions interact with the plasma via the electric field (here $\vec{E}$ ($\vec{E} \parallel ox \parallel \vec{k}_j$; $\vec{k}_j$ — is the wave vector of the $j$-th mode).

Beam–plasma interaction can be described in terms of quasi hydrodynamic equations (Mikhailovsky, 1970):

$$E = -\frac{\partial \varphi}{\partial x}; \ \frac{\partial E}{\partial x} = 4\pi e \left( \overline{\rho}_e - \rho_s - \rho_i \right);$$

$$\frac{\partial V_i}{\partial t} + V_i \frac{\partial V_i}{\partial x} = -\frac{e}{M} E - \nu_{ef} V_i; \frac{\partial \rho_i}{\partial t} + \frac{\partial}{\partial x} \left( \rho_i V_i \right) = 0;$$

$$\frac{\partial \overline{V}_s}{\partial t} + \overline{V}_s \frac{\partial \overline{V}_s}{\partial x} = -\frac{e}{M_o \gamma} \left[ E + c^{-2} \overline{V}_s^2 E \right]; \quad (1)$$

$$\frac{\partial}{\partial t} \left( \gamma \overline{\rho}_s \right) + \frac{\partial}{\partial x} \left( \gamma \overline{\rho}_s \overline{V}_s \right) = 0,$$

where $e$, $Mo$ and $M$ are the electron charge, the rest mass of a beam ion, and the mass of a plasma ion, respectively; $\overline{\rho}_e = N_0 \exp\left( e\varphi / kT_e \right)$; $\varphi$ is the electric potential; $\overline{\rho}_b = N_{0b} + \rho_b$; $\overline{V}_b = V_0 + V_b$; $\rho_b$, $V_b$, $\rho_i$, $V_i$ are deviations of the beam ion density, beam ion velocity, plasma ion density, and plasma ion velocity from their equilibrium values $N_{0b}$, $V_0$, $N_0$, and 0, respectively; and $\gamma = \left( 1 - \overline{V}_s^2 c^{-2} \right)^{-1/2}$. Linearizing Eq. (1) with respect to perturbed quantities, which are assumed to vary in space and time as $\sim exp(i\omega t – ikx)$, we obtain the dispersion relation for the beam–plasma system:

$$1 - \frac{\omega_{0i}^2}{\omega} - \frac{\omega_{0i}^2}{c_s^2 k^2} - \frac{\omega_{0s}^2}{\left( \omega - kV_0 \right)^2 \left( 1 - \frac{\omega - ck}{3ck} \right)} = 0 \quad , \quad (2)$$

$\omega_{0s}^2 = 4\pi e^2 N_{0s} M_0^{-1}$, $\omega_{0i}^2 = 4\pi e^2 N_0 M^{-1}$; $\omega$ is the circular frequency, and k- is

the wavenumber. With $V_0/c_s \gg 1$, $N_{0b}/N_0 \ll 1$ from (2) the approximate dispersion equations

were obtained:

$$\omega_1 \equiv \Omega \approx c_s k_1 \equiv c_s m q; \quad (3)$$

$$\omega_{3,2} - k_{3,2}\, V_0 \approx \mp\, \omega_{0s} + \delta \; ; \frac{\delta}{\omega_{0s}} << 1 \; . \tag{4}$$

Dispersion relation (3) describes an ion sound wave with a positive energy, while dispersion relations (4) describe a slow ($\omega_3$) and a fast ($\omega_2$) beam wave having a negative and a positive energy, respectively. It is easy to see that the slow beam wave ($\omega_3$, $k_3$), fast beam wave ($\omega_2$, $k_2$), and sound wave ($\Omega$, $q$) satisfy the synchronism conditions (Tsytovich, 1967). Taking into account Eqs. (3) and (4), we find from the synchronism conditions that

$$mq \approx 2\,\omega_{0s}\, V_0^{-1} \; . \tag{5}$$

Since ion sound dispersion is weak, the following cascade process is possible:

$$mq + mq \rightarrow 2mq + mq \rightarrow 3mq + mq \ldots mnq \; .$$

Then, after decomposing nonlinear terms in (1) up to the quadratic terms and after using the standard procedure in the weak turbulence (Weiland & Wilhelmsson 1977; Tsytovich 1970), the shortened equations for the complex amplitudes of coupling modes are obtained: beam modes $a_j(j = 1, 2)$ and ion sound $b_k(k = 1, 2, ...mn)$. The analysis of interaction coefficients showed that the systems of such equations describe the stabilized "explosion" (Fainshtein, 1976).

It is also shown that the increment of the growth of ion sound in this case considerably exceeds the values obtained in Fomichev & Fainshtein (1981). Therefore, the mechanism in question occurs much more effectively.

The generable sound is scattered over the fast protons, which move with a speed of $V \sim V_0 \sim 10^{10}$ cm s$^{-1}$ and, according to the mechanism described in Fomichev and Faishtein (1981), the radiation from the source frequency is $\omega_t \approx mqnV$, and frequency separation between the stripes is $\delta\omega_t = mqV$. Taking into account equation (3) and selected parameters ($N_0 \sim 5 \cdot 10^9$ cm$^{-3}$, $N_s \cdot N_0^{-1} \sim 10^{-3}$, the constant magnetic field $\sim 30$ G) we obtain for the emission frequency of $\geq 634$ MHz the value of coefficient $m = 15$, ($7 \cdot 10^2 \leq mn << 15 \cdot 10^3$) and the frequency separation between the adjacent stripes $\delta\omega_t \approx 15$ MHz. The obtained value of $\delta\omega_t$ corresponds to the observed frequency separation in the decimeter wave band (Chernov, 2006). In the given above estimations, the wavelength of the ion sound $\sim 100$ m, the initial frequency $\sim 1.0$ kHz and the cyclic frequency of the slow beam wave $\omega_3 \sim 7 \cdot 10^2\,\omega_{0i}$ (which correspond to $\sim 1$ GHz).

Discrete emission bands are possible when the width of each emission band will be less than the value of the frequency separation. This condition imposes a restriction on the dispersion of the beam velocities. As shown in such estimations, and implemented in Fomichev and Fainshtein (1981), the beam of protons must be sufficiently quasi-monoenergetic, in the case in question, $\Delta V_0 / V_0 < 10^{-3}$.

The DPR mechanism was even used to interpret zebra-like stripes in microwave radiation of the Crab Nebula pulsar, despite numerous problems with the plasma parameters (Zheleznyakov et al., 2012). Those problems motivated Karlický to develop a ZP model, based on the FMA waves, for pulsars.

The discussed explosive instability could be a more probable model for the interpretation of zebra stripes in the radio emission of a pulsar in the Crab nebula under extreme physical parameters peculiar to pulsars than the DPR model because it does not depend on the ratio of plasma and gyrofrequency.

## 4.2. Other models

From other zebra models, the formation of transparency and opacity in bands during the propagation of radio waves through regular coronal inhomogeneities is the most natural and promising mechanism (Laptuhov, Chernov, 2012). It explains all main parameters of the regular ZP. The dynamics of the ZP stripes (variations in the frequency drift, stripe breaks, etc.) can be associated with the propagation of inhomogeneities, and their evolution and disappearance. Inhomogeneities are always present in the solar corona, however direct evidences of the existence of inhomogeneities with scales of several meters in the corona are absent, although ion-sound waves could serve this purpose.

The model of a nonlinear periodic space charge waves in plasma (Kovalev, 2009) is also a very natural mechanism in solar flare plasma. However, in the case of intrinsic plasma emission, it gives a constant frequency separation between stripes of $\approx \omega_B$, while the observations verify the increase in the frequency separation with frequency. In addition, the condition of achieving strong nonlinearity remains uncertain.

The mechanism scattering fast protons on ion-sound harmonics in explosive instability looks very uncommon, and it requires a number of strict conditions (Fomichev et al. 2009), although fast protons always exist in large flares, and the presence of nonisothermic plasma is completely feasible in the shock wave fronts.

The last two models could be useful in describing large radio bursts. All three models are related to a compact radio source. The number of discrete harmonics does not depend on the ratio of plasma frequency to gyrofrequency in the development of all three models. The latter circumstance can eliminate all the difficulties that arise in the DPR model.

## 5. Conclusion

We discussed the difficulties of the DPR mechanism according to the latest publications and showed the potential of some promising alternative mechanisms.

First of all, a radio source of a ZS is usually located in magnetic islands after CME ejections. Therefore, a close connection of the ZS with fiber bursts is simply explained by the acceleration of fast particles in regions of magnetic reconnection in the bottom and the top of magnetic islands.

A wavelike or saw-tooth frequency drift of stripes was explained by the switching of whistler instability from the normal Doppler cyclotron resonance into the anomalous one (Fig. 9). Such switching must lead to a synchronous change in the frequency drift of stripes and the spatial drift of the radio source, because the whistlers generated at normal and anomalous resonance move in opposite directions. New injections of fast particles cause a sharp change in frequency drift of stripes in instantaneous columns of a ZS. Low frequency absorptions (black stripes in a ZS) are explained by attenuation of plasma wave instability due to diffusion of fast particles by whistlers.

The superfine structure is generated by a pulsating regime of the whistler instability with ion-sound waves (Chernov et al. 2003). Rope-like chains of fiber bursts are explained by the periodic whistler instability between two fast shock fronts in a magnetic reconnection region (Chernov, 2008). In the whistler model, zebra stripes can convert into fiber bursts and inversely, as shown in Figures 2 and 3.

Separately, we isolated the prospects for explosive instability, which can completely work in the major events that have the ejections of protons.

For a comprehensive discussion of comparative analysis of observations of the ZS and fiber bursts, and different theoretical models we refer the reader to the reviews of Chernov (2012; 2016) (freely available at http://www.izmiran.ru/~gchernov/).

**Acknowledgements**

The authors are grateful to Chinese colleagues based at National Astronomical Observatories, Chinese Academy of Science, including Yihua Yan, Chenming Tan and Guannan Gao. This work is also supported by the Russian Foundation for Basic Research under Grant 17-02-00308.